%% file: main.tex
\documentclass[sigconf]{acmart}

\setcopyright{none} 
\acmConference[Anonymous Submission to ACM CCS 2020]{ACM Conference on Computer and Communications Security}{Due November 11-15, 2019}{London, UK}
\acmYear{2019}

\usepackage{array}
\usepackage{url}
\usepackage{cleveref}
\usepackage{amsmath,amssymb,amsfonts}
\usepackage{algorithmic}
\usepackage{graphicx}
\usepackage{textcomp}
\usepackage{xcolor}
\usepackage{booktabs}

\usepackage{caption}
\usepackage{subcaption}
\usepackage{float}
\usepackage{textcomp}
\usepackage{multirow}

\newcommand{\post}{p}
\newcommand{\Post}{\mathit{P}}
\newcommand{\hashtag}{h}
\newcommand{\Hashtag}{\mathit{H}}
\newcommand{\capt}{t}
\newcommand{\Capt}{\mathit{T}}
\newcommand{\img}{i}
\newcommand{\Img}{\mathit{I}}
\newcommand{\user}{u}
\newcommand{\User}{\mathit{U}}
\newcommand{\pair}{\{u, v\}}
\newcommand{\Pair}{\mathit{V}}
\newcommand{\loc}{\ell}
\newcommand{\Loc}{\mathit{L}}
\newcommand{\RelVar}{\mathcal{R}_{\pair}}
\newcommand{\OSN}{\mathcal{G}}
\newcommand{\partOSN}{\mathcal{G}'}
\newcommand{\Edge}{E}
\newcommand{\data}{\mathit{D}}

\newcommand{\HA}{\mathtt{HA}}
\newcommand{\CA}{\mathtt{TA}}
\newcommand{\IA}{\mathtt{IA}}
\newcommand{\LA}{\mathtt{LA}}
\newcommand{\NA}{\mathtt{NA}}
\newcommand{\SA}{\mathtt{SA}}
\newcommand{\HolA}{\mathcal{M}}
\usepackage{setspace}
\usepackage{xargs}     
\usepackage{xcolor}  
\usepackage[disable,colorinlistoftodos,textsize=tiny]{todonotes}

\begin{document}
\title{\Large \bf Everything About You: A Multimodal Approach towards 
Friendship Inference in Online Social Networks}
\author{Tahleen Rahman}
\affiliation{\institution{CISPA Helmholtz Center for Information Security}}
\author{Mario Fritz}
\affiliation{\institution{CISPA Helmholtz Center for Information Security}}
\author{Michael Backes}
\affiliation{\institution{CISPA Helmholtz Center for Information Security}}
\author{Yang Zhang}
\affiliation{\institution{CISPA Helmholtz Center for Information Security}}

\begin{abstract}
Most previous works in privacy of Online Social Networks (OSN) focus on a restricted scenario of using one type of information to infer another type of information or using only static profile data such as username, profile picture or home location. However the multimedia footprints of users has become extremely diverse nowadays. In reality, an adversary would exploit all types of information obtainable over time, to achieve its goal. In this paper, we analyse OSN privacy by jointly exploiting longterm multimodal information. We focus in particular on inference of social relationships. We consider five popular components of posts shared by users, namely images, hashtags, captions, geo-locations and published friendships. Large scale evaluation on a real-world OSN dataset shows that while our monomodal attacks achieve strong predictions, our multimodal attack leads to a stronger performance with AUC (area under the ROC curve) above 0.9. Our results highlight the need for multimodal obfuscation approaches towards protecting privacy in an era where multimedia footprints of users get increasingly diverse.
\end{abstract}


\maketitle

\input{introduction}
\input{model}

\input{dataset}

\input{captions}
\input{hashtags}

\input{images}
\input{location}
\input{holisticattack}
\input{discussion}

\input{related_work}
\input{conclusion}

\bibliographystyle{ACM-Reference-Format}
\bibliography{biblio}
\newpage
\input{appendix}

\end{document}

%% file: introduction.tex
\section{Introduction}
\label{sec:intro}
\begin{figure}[t]
\centering
\includegraphics[width=\columnwidth]{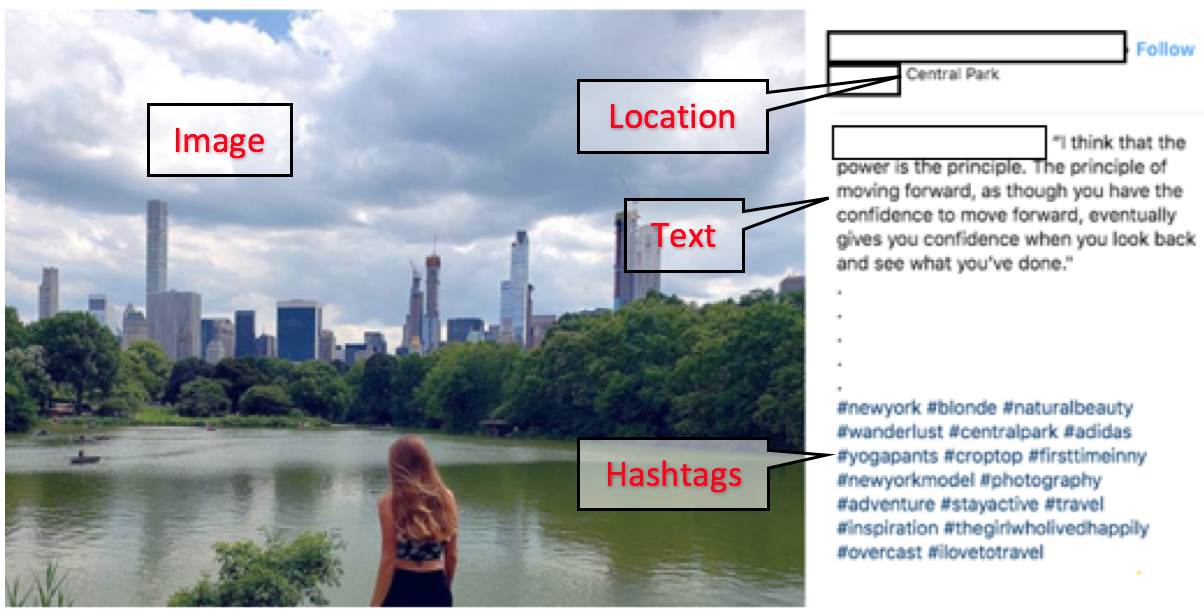}
\caption{Example of an Instagram post containing multimodal information. We investigate privacy threats emerging in such a multimodal setting against an adversary that uses a combination of images, locations, text and hashtags shared with a post as well as already revealed friendships between users of an OSN. \label{example}}
\end{figure}

Online Social Networks (OSNs) have become an indispensable part of people's life in the modern society.
Leading players in the businesses, such as Facebook and Instagram, have acquired a large number of users.
Initially, people used OSNs mostly to articulate their friendships.
With time, online activities have become increasingly elaborate and diverse.
Users publish tweets on Twitter, statuses on Facebook and captions on Instagram.
Photos constitute a large proportion of the social media content.
People also share their locations, popularly known as check-ins, via GPS-enabled smartphones. 
More recently, hashtags have become very popular for content discovery and advertisement. Figure~\ref{example} shows an example of a typical OSN/Instagram post 
that consists of a photo, shared with a textual caption, 
supplemented with hashtags, and a geo-tagged location. 

Extensive amounts of such diverse user data is generated in OSNs everyday which has been exploited by various parties to gain a deeper understanding of our society.
On the downside, privacy risks stemming from data sharing in OSNs have become a major concern.
Researchers have highlighted potential attacks on various aspects, e.g., social security number prediction~\cite{Acquistipnas}, stalking and re-identification~\cite{gross2005information} and various inference attacks.
However, most prior works in this direction explore only a single kind of user generated content to infer another hidden attribute.
For example, Backes et al.~\cite{walk2frns} infer social relations from users' locations; Zhang et al. \cite{tagvisor} infer user locations from hashtags; Shoshitaishvili et al.~\cite{shoshitaishvili2015portrait} infer  pairs that are dating using facial recognition.
In a realistic scenario, adversaries would utilize not just a single kind but as many different kinds of data as possible to achieve their goal. 
Works that do consider heterogeneous data \cite{goga2015reliability} utilize only a single instance of static profile data such as username, profile picture, home location, etc or a footprint of only one modality \cite{liu2014hydra}. 
However, users generate much more than only one kind of content and they generate this multimodal content continuously. 
The multimodal footprint of users built up over time from the data they post everyday, collectively gives a much more faithful reflection of the user's interests.
Such multimedia footprints of users are also getting increasingly diverse.

In order to fully assess the privacy risks arising from such multimodal footprints over time, 
a comprehensive evaluation is needed which, to our knowledge,
has not been addressed in the literature.
This paper aims to fill this gap with a framework incorporating multimodal data in OSNs to assess users' privacy risks.

Out of the various privacy risks in OSNs, 
we concentrate on inferring social relations.  
Social relationships are recognized to pose high threats to user privacy.
In addition to revealing users' social identities, social relationships can potentially leak further sensitive attributes about a user, which the user explicitly intends to hide in their own profile. In practice, OSN users have also begun to increasingly hide their social circles~\cite{DJR12}.

\subsection{Contributions}
We conduct the first large-scale privacy assessment 
of collective multimodal information shared continuously over time in OSNs against an adversary 
that leverages not just a single component
but combines all available components of user-generated content.
In particular, we consider five kinds of information: images, texts, location, hashtag, and (incomplete) social relations.
We evaluate on a real-life dataset containing 22 million user posts in two cities which we collected from Instagram.

Our contributions are two-fold.
First, we introduce three novel monomodal friendship inference attacks
based on hashtags, images, and text, respectively.
To the best of our knowledge, 
this is the first work that uses text and image modalities to conduct friendship inference in social networks. 
Our hashtag attack massively outperforms the only prior work on friendship inference from hashtags \cite{zhang2019language} by 7\%.
We use self-engineered statistical features, such as the entropy and frequency of usage patterns for our hashtag and text based attacks.
For images, we rely on a state-of-the-art ResNet model to extract information from images.
Experiments show that our hashtag and text attacks achieve strong performance with AUCs above 0.86 and 0.83 respectively. 
Our image attack, while being the weakest, still performs well with AUCs up to 0.637.
 
Second, we design a multimodal attack that combines 
five components of posts to infer social relations.
Extensive experiments show that this attack outperforms all monomodal attacks with AUCs above 0.9.
We also show that the multimodal attack exploits post components that compliment each other 
and the success of most attacks increases significantly when combined with additional components. 
We further demonstrate the robustness of the multimodal attack against information hiding, and show that even after deleting 50\% of the users' posts,
the drop in the attack performance is relatively low.

\subsection{Organization}
In~\cref{sec:model}, we present our user and adversary model and notations.
In~\cref{subsec:dataset}, we introduce our dataset.  
We present the design and experimental evaluation of our attacks using texts, hashtags, images and locations together with social network edges in sections~\ref{captions},~\ref{hashtags},~\ref{images} and ~\ref{locations} respectively.
In~\cref{holistic}, we finally present the design and evaluation for our multimodal attack.
We discuss limitations of the work and potential directions for future work in~\cref{discussion}
We review the literature in~\cref{sec:relwork}, and finally conclude the paper in~\cref{sec:conclu}.

%% file: model.tex
\section{Model}
\label{sec:model}

\subsection{User Model}
We define an OSN by an undirected, unweighted graph $\OSN = (\User, \Edge)$, 
where the set $\User$ contains all the users 
and $\Edge \subseteq \{ \pair  \vert   u \in \User, v \in \User, u \neq v\}$ is the set of friendships.
An edge exists between 2 users if they mutually and visibly declare each other as friends in their profiles.

A user $\user$ shares a set of posts $\Post_\user $ which can be composed of images, texts, hashtags and location check-ins. We define a single post $\post_\user \in \Post_\user$ as a 4-tuple $\langle \img_\post$, $\loc_\post$, $\capt_\post$, $\Hashtag_\post \rangle$ which denotes that a user $\user \in \User$ posts an image $\img_\post$, at location $\loc_\post $ (checks in at $\loc$), with a text $\capt_\post $ annotated with the set of hashtags 
$\Hashtag_\post $.
All the posts of  $\user$ is defined by $\Post_\user = \{\post_\user^1, \post_\user^2  \ldots \post_{\user}^{|\Post_\user|}\}$.

We denote the overall set of posts by $\Post$, images  by $\Img$, textual posts by $\Capt$, the set of  hashtags by $\Hashtag$ and the set of locations $\Loc$. 
Accordingly, for $\user$, $\Img_\user \subset \Img $ represents the set of images, $\Hashtag_\user \subset \Hashtag$ represents the set of hashtags, $\Capt_\user \subset \Capt$ the set of texts and $\Loc_\user \subset \Loc$ the set of check-ins shared by $u$.

\subsection{Adversary Model}
We consider an adversary that tries to predict whether a pair of individuals $u$ and $v$ that explicitly hide their friendship are socially related or not.  
A pair of users can hide their friendship in different ways depending on the OSN application, for example by not following each other or not enlisting each other in their respective friend lists. Users may want to do so for several reasons such as suppressing a workplace romance, or protecting themselves from \emph{target development} by surveillance agencies (\cite{NSA} reveals how NSA collects mobility and travel habit data to spot new unknown associates from already known targets).
The adversary could be any party with the ability to crawl an OSN and access user posts.

The adversary has access to posts shared by the pair of users $\pair$ namely $\Post_\user $ and $\Post_\user $ as well as the friendships explicitly published on the OSN, $\Edge $.
A feature $\vec{x}^{\data}_{\pair} $ for a pair $\pair$, is a function of the posts $\Post_u$ and $\Post_v$ shared by them and friendships published on the OSN, $\OSN$.  
The adversary constructs features $\vec{x}^{\data}_{\pair} $ from each component of posts: hashtags, location check-ins, texts and images as well as published friendships such that $D \subseteq \{\Hashtag, \Capt, \Img, \Loc , \Edge \}$.
The prediction attack is then a binary classification task $f:\vec{x}^{\data}_{\pair} \rightarrow \{0, 1\}$, where we encode the existence of friendship by the positive class and the absence of friendship by the negative class.
In the upcoming sections, we describe the process of constructing the features $\vec{x}^{\data}_{\pair} $ from each of the individual data components. 

We distinguish between 2 kinds of adversaries for ease of readability. 
The first kind of adversary has access to and utilizes only a single component of OSN posts.
We name them according to the information they have access to/use:

1. The Text Adversary, $\CA$ which uses the text component of OSN posts. We denote the feature vector used for the corresponding prediction task by $\vec{x}^{\Capt}_{\pair}$.

2. The Hashtag Adversary,  $\HA$ which uses the hashtag component of OSN posts. We denote the corresponding  feature vector by $\vec{x}^{\Hashtag}_{\pair}$.

3. The Image Adversary,  $\IA$ which uses the image component of OSN posts. We denote the corresponding  feature vector by $\vec{x}^{\Img}_{\pair}$.
4. The Location Adversary, $\LA$ which uses the location component of OSN posts. We denote the corresponding  feature vector by $\vec{x}^{\Loc}_{\pair}$.

5. The Network Adversary,  $\NA$ which uses the friendships explicitly published by OSN users. Since friendships on the OSN are represented by edges $\Edge$, we denote the corresponding  feature vector by $\vec{x}^{\Edge}_{\pair}$.

We refer to this group of attacks as \emph{monomodal attacks}.

The second kind of adversary combines multiple information components shared by users of OSNs. We name such an adversary as the Multimodal Adversary $\HolA$ and this group of attacks as \emph{multimodal attacks}. When the $\HolA$ uses a subset $D'$ of post components, we call it a Subset Adversary $\SA^{D'}$,  $D' \subset \{ H, C, I, L, E  \}$ and $ 2 \leq |D'| \leq 4$.

%% file: dataset.tex
\section{Dataset}
\label{subsec:dataset}
To evaluate our attacks, we use a dataset of Instagram, one of the largest OSNs \footnote{\url{https://www.dreamgrow.com/top-15-most-popular-social-networking-sites/}} of this era, which has been used previously for other works ~\cite{walk2frns}, \cite{tagvisor}, \cite{rahman2019fairwalk}, \cite{zhang2019language}.
Ideal as it might be, it remains infeasible to obtain the ground truth for a dataset where people hide their friendship online in spite of secretly being friends in real life. Therefore we adopt the best possible alternative and evaluate our attacks on a simulation of the ideal dataset by using a real life Instagram dataset.
Compared to other OSNs like Twitter which is mainly used for sharing news, current events and public statements, Instagram is better suited for friendship inference as it is mainly used to share topics related to lifestyle, interests, hobbies, leisure and travel etc. Consequently, Instagram friends tend to be socially closer in real life as well. On the other hand, crawling Facebook (the largest OSN) is prohibited.
Moreover, Instagram focuses more on sharing of visual content and hashtags and is used by many brands to humanize their content for better connecting customers and growing brand awareness.
A typical Instagram post as shown in Figure~\ref{example} consists of a photo(s) or video along with captions and hashtags and is geo-tagged with locations or Points of Interest (POIs).

The dataset contains 8,309,482 posts from 10430 users in Los Angeles and 14,671,556 posts from users in New York comprising of photos, captions, hashtags and tagged locations that were publicly posted along with the corresponding user ID, and for each user, the respective followers. 
Moreover, there are 18,221 friend pairs for LA out of which we obtain 17,043 pairs that have at least 1 out of the 5 types of feature data.
From NY we collected 41,501 friend pairs, out of which 40,883 pairs have feature data in at least 1 out of the 5 attack groups. 
For individual attacks, we eliminate friend pairs that do not have the corresponding feature data.

We further do not target friendship inference between Instagram accounts that are likely to be celebrities or bots.  
Therefore, we filter out accounts whose numbers of followers are above the 90th percentile (celebrities) or below the 10th percentile (bots) following prior works, \cite{cho2011friendship, tagvisor,scellato2011exploiting, walk2frns}.

\smallskip
\noindent\textbf{Ethics}. Our institution neither provides an IRB nor mandates (or enables) approval for such experiments.
Nevertheless, only publicly available data is collected via Instagram's public API in 2016. We anonymize the datasets by removing all users' screen names, and replacing their Instagram IDs with randomly generated numbers. Only these anonymized datasets are then stored for experiments in a central server with up-to-date software, encrypted hard drives and access only possible from two white-listed internal IP addresses using authorized SSH
keys.

%% file: captions.tex
\section{Attack using Text}
\label{captions}

In this section, we describe our friendship inference attack for the Text Adversary $\CA$ followed by an experimental evaluation.

\subsection{Attack Methodology}
\label{captionattack}
The Text Adversary $\CA$ has access to the text like tweets, statuses or captions shared by users. 

 We design features for this attack in two phases:
 
\subsubsection{Phase I}

We first collect all textual data shared by each user and tokenize the text sequences. 
We weigh the tokens according to the number of users in the overall OSN that use them such that tokens used by a majority of users have lower weight than those used by fewer users. 
To this end we use a Bag of Words model with \textit{TF-IDF} (Term-Frequency Inverse Document-Frequency).
For a term $t$ and a document $d$, $\textit{TF-IDF(t,d)}$ is the product of the term frequency $\text{tf(t,d)}$ and inverse document frequency  $\text{idf}(t)$.
\[
\textit{TF-IDF(t,d)}=\textit{tf(t,d)} \times \textit{idf(t)}
\]
where term frequency is defined as
$\textit{tf(t,d)} = \frac{n_{t,d}}{\sum_{t' \in d}{n_{t',d}}}$
with $n_{t,d}$ as the number of times term $t$ appears in document $d$ and $\sum_{t' \in d}{n_{t',d}}$ as the total number of terms in $d$.
Meanwhile, inverse document frequency is
$\textit{idf}(t) = log{\frac{n_d}{1+\text{df}(d,t)}}$
with $n_d$ as the total number of documents and $\text{df}(d,t)$ as the number of documents that contain term $t$.
As we see, the IDF reweights the popular but less interesting words.

The resulting \textit{TF-IDF} vectors are then normalized by the Euclidean norm:
\[
v_{norm} = \frac{v}{||v||_2} = \frac{v}{\sqrt{v{_1}^2 +
v{_2}^2 + \dots + v{_n}^2}}.
\]
We therefore convert the text sequences of all users into a matrix of \textit{TF-IDF} vectors. 

Assuming $W$ is the overall set of words in the textual dataset for a city or in other words,
 the vocabulary of the Text Adversary $\CA$, 
 we denote each user's \textit{TF-IDF} vector by $\vec{\mathit{T}_\user} \in R^{|W|} $ such that 
 $ \vec{\mathit{T}_\user} =( \textit{TF-IDF}(t_{1}, u)  $,  
 $\textit{TF-IDF}(t_2,u),  \ldots , \textit{TF-IDF}(t_{|W|},u) )  $.

\subsubsection{Phase II}
For each user pair $\pair$, we compare the \textit{TF-IDF} vectors $\vec{\mathit{T}_u}$ and $\vec{\mathit{T}_v}$. 
We calculate 8 common pairwise distances to form the feature vector for the Text Adversary namely 
$\vec{x}^{\Capt}_{\pair}  =
(\mathrm{cosine}(\vec{\mathit{T}_u},\vec{\mathit{T}_v}), \:
\mathrm{  euclidean}(\vec{\mathit{T}_u},\vec{\mathit{T}_v}),  \:
\mathrm{ correlation}(\vec{\mathit{T}_u},\vec{\mathit{T}_v}), 
$ \\ $
\mathrm{ chebyshev}(\vec{\mathit{T}_u},\vec{\mathit{T}_v}), \:
\mathrm{ bray\_curtis}(\vec{\mathit{T}_u},\vec{\mathit{T}_v}),  \:
\mathrm{  canberra}(\vec{\mathit{T}_u},\vec{\mathit{T}_v}),
$ \\ $
\mathrm{  manhattan}(\vec{\mathit{T}_u},\vec{\mathit{T}_v}), \:
\mathrm{  sq\_Euclidean}(\vec{\mathit{T}_u}, \vec{\mathit{T}_v}) 
)$.
Table \ref{dist_}  contains the definitions of each measure.

\begin{table}[!ht]
\centering
\caption{Definitions of the common pairwise distance measures used for the textual attack.
$\overline{\vec{\mathit{T}_u}}$ represents the mean of $\vec{\mathit{T}_u}$.
 $\vec{\mathit{T}_u}_i$ represents the $i$th element in $\vec{\mathit{T}_u}$. 
 \label{dist_}}
\begin{tabular}{l|c}
\toprule
measure & definition \\ [2pt]
\midrule
$\mathrm{  cosine}(\vec{\mathit{T}_u},\vec{\mathit{T}_v}) $ & $\frac{\vec{\mathit{T}_u}\cdot\vec{\mathit{T}_v}}{\vert \vert\vec{\mathit{T}_u}\vert\vert_2\ \vert \vert\vec{\mathit{T}_v}\vert\vert_2} $\\ [9pt]
$\mathrm{  euclidean}(\vec{\mathit{T}_u},\vec{\mathit{T}_v})$ & $ \vert \vert\vec{\mathit{T}_u}-\vec{\mathit{T}_v}\vert\vert_2 $ \\  [5pt]
$\mathrm{  correlation}(\vec{\mathit{T}_u},\vec{\mathit{T}_v})$ & $ \frac{(\vec{\mathit{T}_u}-\overline{\vec{\mathit{T}_u}})\cdot(\vec{\mathit{T}_v}-\overline{\vec{\mathit{T}_v}})}{\vert \vert\vec{\mathit{T}_u}-\overline{\vec{\mathit{T}_u}}\vert\vert_2\ \vert \vert\vec{\mathit{T}_v}-\overline{\vec{\mathit{T}_v}}\vert\vert_2}$ \\ [12pt]
$\mathrm{  chebyshev}(\vec{\mathit{T}_u},\vec{\mathit{T}_v})$ & $ \max_{i= 1}^{d}\vert \vec{\mathit{T}_u}_i-\vec{\mathit{T}_v}_i\vert$ \\ [6pt]
$\mathrm{  bray\_curtis}(\vec{\mathit{T}_u},\vec{\mathit{T}_v})$ & $ \frac{\sum_{i= 1}^{d}\vert \vec{\mathit{T}_u}_i-\vec{\mathit{T}_v}_i\vert}{\sum_{i= 1}^{d}\vert \vec{\mathit{T}_u}_i+\vec{\mathit{T}_v}_i\vert}$ \\ [10pt]
$\mathrm{  canberra}(\vec{\mathit{T}_u},\vec{\mathit{T}_v})$ & $ \sum_{i= 1}^{d}\frac{\vert \vec{\mathit{T}_u}_i-\vec{\mathit{T}_v}_i\vert}{\vert \vec{\mathit{T}_u}_i\vert + \vert\vec{\mathit{T}_v}_i\vert}$ \\ [8pt]
$\mathrm{  manhattan}(\vec{\mathit{T}_u},\vec{\mathit{T}_v}) $ & $ \sum_{i= 1}^{d}\vert \vec{\mathit{T}_u}_i-\vec{\mathit{T}_v}_i\vert$ \\  [6pt]
$\mathrm{ sq\_Euclidean}(\vec{\mathit{T}_u},\vec{\mathit{T}_v})$ & $ \vert \vert\vec{\mathit{T}_u}-\vec{\mathit{T}_v}\vert\vert_2^2 $ \\  [5.5pt]  
\bottomrule
\end{tabular} 
\end{table}

\subsection{Evaluation}
\label{cap_data}
We now present the preprocessing performed on the raw dataset, followed by the experimental set-up and the metric we use for evaluating the  prediction performance. Finally we present the results of the experiments.

\subsubsection{Data preprocessing}

In LA, we have 10,018 users that share a total of 7,146,810 captions with their Instagram posts.
For NY we have 12,756,384 captions shared by 18,713 users.
In order to remove terms that are too rare and too frequent, we filter out words that appear in caption sequences of more than 100 users and less than 2 users.
Following this step, we obtain 16,878 out of a total of 17,043 friend pairs from LA and 40,228 out of 40,883 friend pairs from NY.
We randomly sample an equal number of stranger pairs.
Table \ref{cap_pro_dataset} shows the number of users, terms and friend pairs that we have after pre-processing.

\begin{table}[!ht]
\centering
\caption{Pre-processed caption data statistics. \label{cap_pro_dataset}}
\scalebox{0.9}{
\begin{tabular}{lcc}
\toprule
 & LA & NY \\ 
\midrule
 users & 10,018 & 18,713 \\ 
 terms & 587,190 & 886,120 \\ 
friend pairs & 16,878 & 40,228 \\ 
\bottomrule
\end{tabular} 
}
\end{table}

\subsubsection{Experimental Set-up} 
We rely on a random forest with 100 trees to carry out this and all further attacks as it outperforms other classifiers like Gradient Boosting Machine, Ada Boost, Support Vector Machine and Logistic Regression in our experiments.
Increasing the number of trees beyond 100 did not significantly change our performance. 

We perform a 5-fold cross validation with a 80-20 split into training and test set.
We do not tune any parameters and therefore do not require a validation subset.

\subsubsection{Metric}
We evaluate our attack using the AUC (Area Under the receiver operating characteristic (ROC) Curve) metric.
AUC exhibits a number of desirable properties when compared to accuracy such as decision threshold independence and invariance to a priori class probabilities.
The latter is particularly desired for friendship inference in OSNs in order to correctly evaluate the attacks in our down-sampled prediction space.
Moreover, there exists a conventional standard for interpreting AUC in a rather intuitive manner. For AUCs in the range [0.5, 1], with 0.5 associated with random guessing and 1 with perfect guessing, values above 0.8 represent a very good prediction and values above 0.9 represent excellent prediction. This simplicity renders comparison against baseline models dispensable. 

\begin{figure}[t]
\centering
\includegraphics[width=\columnwidth]{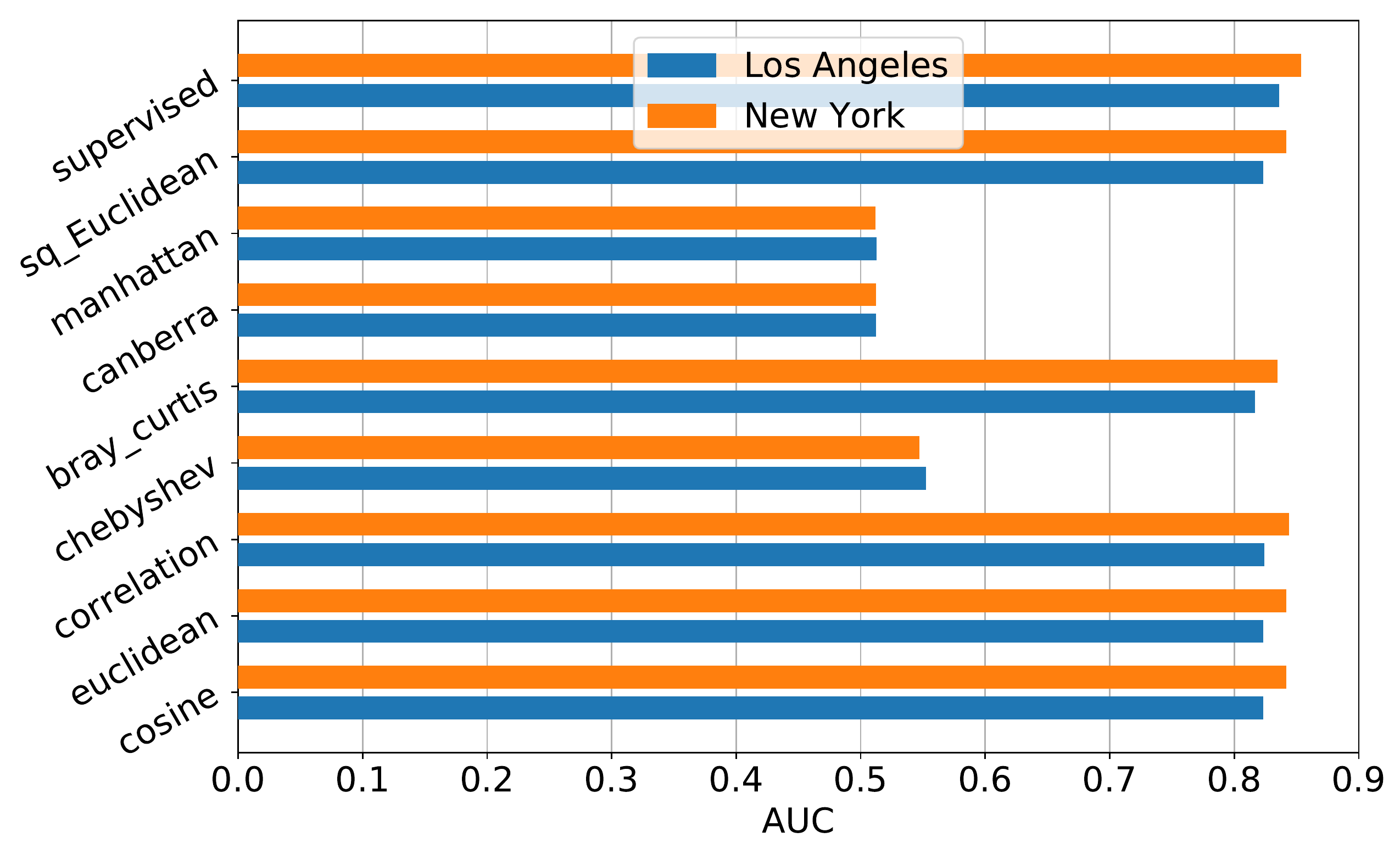}
\caption{Comparison of individual features for our Text Adversary \label{fig:cap_unsup}}
\end{figure}

\subsubsection{Results}
We first assess how each the 8 individual distance measures perform on their own and thus  compare their AUC values in the unsupervised setting in Figure~\ref{fig:cap_unsup}.
We see that correlation achieves the highest score in both cities, with an AUC value of 0.82 for LA and 0.84 for NY. Squared Euclidean, Bray-Curtis, Euclidean and cosine distance achieve only slightly lower AUCs.

This shows that the success of our attack is not dependant on the city from which we collect our data.
In the supervised setting, we further achieve a much stronger performance in both cities with AUC of 0.83 for LA and 0.85 for NY.

It is worth noting that newer, more advanced NLP techniques might be more optimal for the prediction task. However, this paper is focussed on highlighting the privacy risks from an adversary that combines multiple modalities and the need to defend against such multimodal attacks. Other state-of-the-art techniques can be  adopted similarly in a straightforward manner.

%% file: hashtags.tex
\section{Attacks using Hashtags}
\label{hashtags}
We now describe our friendship inference attack for the Hashtag Adversary $\HA$ along with an experimental evaluation.

\subsection{Attack Methodology}\label{hashtag_attacks}

We train a machine learning classifier with the auxiliary knowledge of the adversary which consists of a 10-dimensional feature space which we present below. 
The adversary uses the trained classifier and carries out his inference attack on a target user pair by computing these scores for the pair using their hashtag sharing information.

\begin{figure}[t!]
\centering
\includegraphics[width=0.5\textwidth]{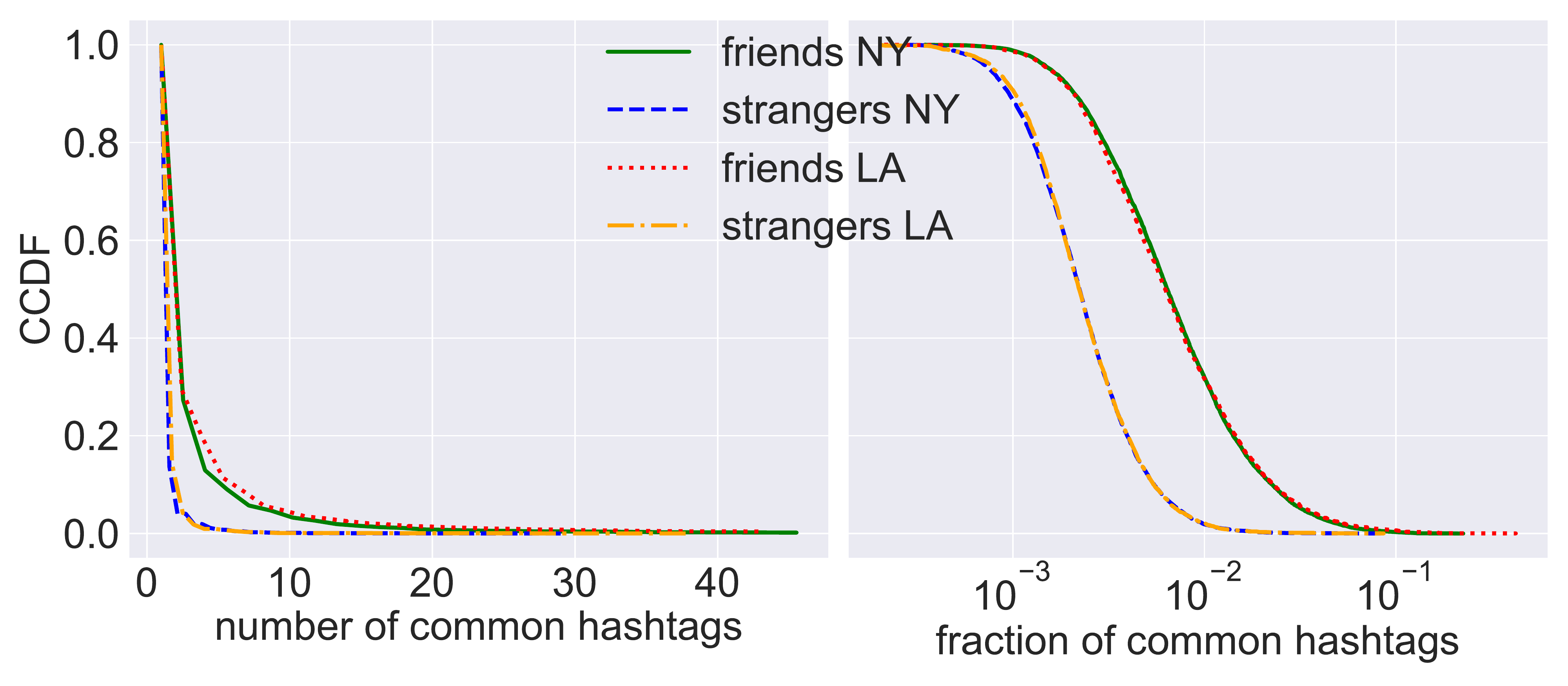}
\caption{CCDF of the number and fraction of hashtags in common between friends versus strangers.
\label{fig:htprops}}
\end{figure}

Our feature design is influenced by certain properties we observed during preliminary analyses on our dataset. These properties fall into 2 main directions namely the hashtags shared in common by a user pair and the information content or the relevance of individual hashtags. 

Firstly in Figure \ref{fig:htprops} we observe that in both cities, friends tend to have a higher number of hashtags in common as compared to strangers. 
The fraction of common hashtags, (number of overlapping hashtags divided by the total number of hashtags shared by both users), also shows a similar property;
a high fraction of the overall hashtags shared by a user overlaps with the hashtags shared by a friend of the user.  
Moreover, we only consider stranger pairs that have at least 1 hashtag in common. Including stranger pairs with no common hashtags would only further increase the gap between the friend and stranger curves.
The number of common hashtags shared by a user pair therefore appears to be a very good indicator of the existence of friendships between them which motivates the design of our first 4 features.
Denoting the set of hashtags shared by user $u$ and user $v$ by $\Hashtag_u $ and $\Hashtag_v$ respectively, we define $\mathit{x_{comm}}$ and $\mathit{x_{frac}}$ as the number and fraction respectively of hashtags shared by both users. 
\begin{align}
\mathit{x_{comm}} & = |\Hashtag_u \cap  \Hashtag_v| \\
\mathit{x_{frac}} & = \frac{|\Hashtag_u \cap  \Hashtag_v|}{|\Hashtag_u \cup  \Hashtag_v|}
\end{align}

We define $\mathit{x_{dotp}}$ as the (dot product of) number of times the common hashtags were shared by both users.  We represent the number of times a particular user $u$ shares each hashtag $h_i \in H$ as a user-hashtag count vector $ \vec{n_{u}}= (n_{u, h_1}, n_{u, h_2}, \ldots n_{u, h_{|H|}})$, where $n_{u, h_i}$ is the number of times $\hashtag_i$ is shared by $u$. We use $\vec{n_v}$ analogously for user $v$ in a pair. We define $\mathit{x_{cos}}$ as the cosine similarity between the user-hashtag count vectors $\vec{n_{u}}$ and $\vec{n_{v}}$ for a user pair $\pair$. Thus,
 
\begin{align}
\mathit{x_{dotpro}} &= \vec{n_{u}} \vec{n_{v}}  \\
\mathit{x_{cosine}} &= \frac{ \vec{n_{u}} \vec{n_{v}} }  {  \sqrt{ \vec{n_{u}}^2      \vec{n_{v}}^2         }  }
\end{align}

 \begin{figure}[t!]
\centering
\includegraphics[width=0.5\textwidth]{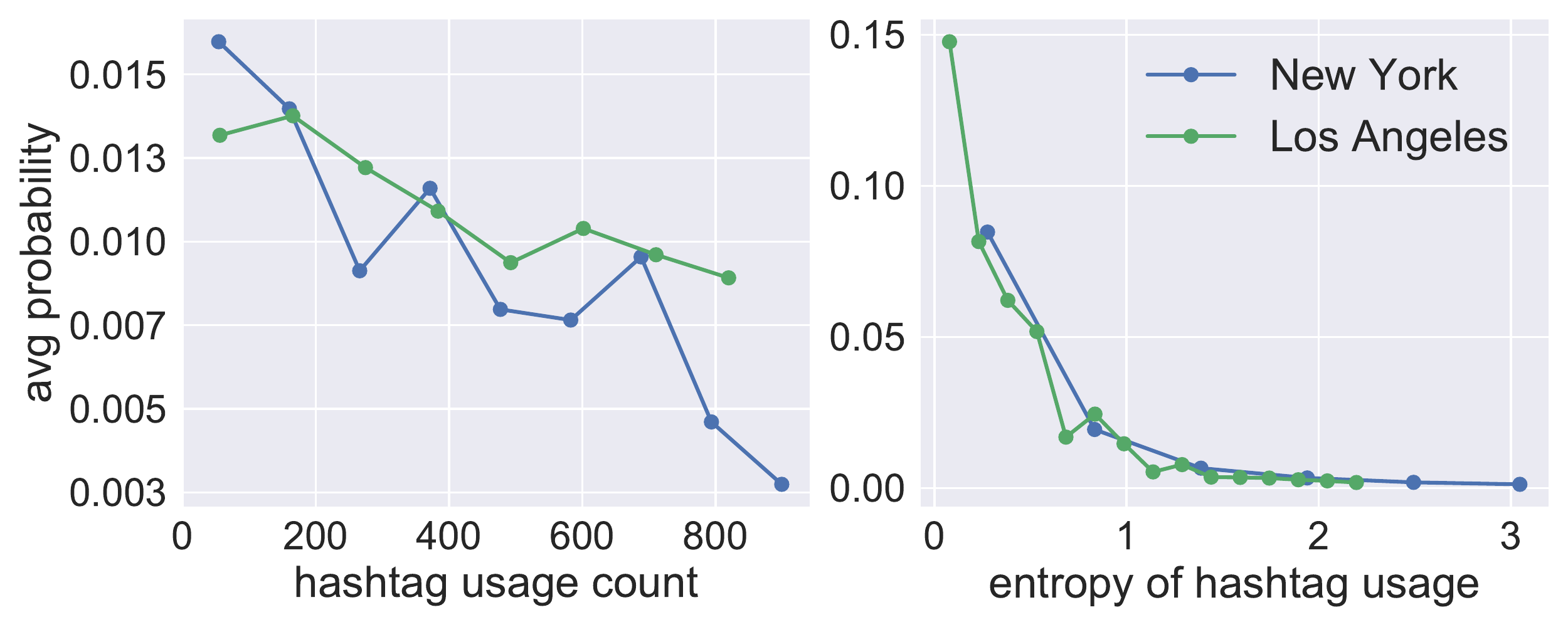}
\caption{Average probability that two users who share the same hashtag are friends as a function
of (left) overall popularity of the hashtag and (right) entropy of the hashtag usage. The left x-axis marks the usage count which is the total number of times the corresponding hashtag was shared by all users in the OSN. The right x-axis marks the entropy of the vector counting the number of times the corresponding hashtag was shared by various users.
\label{fig:htperc}}
\end{figure}

The second property that can be observed in Figure \ref{fig:htperc} (left) is that hashtags shared by only a small number of users regularly are more likely to hold special significance for them, such as hobbies, places visited, private clubs, jargon, acronyms and idioms and so on.
Conversely, hashtags shared at least as frequently but by a larger number of users are likely to be merely common terms and not characteristic of a socially related users.

We therefore formalize the notion that a shared hashtag is more likely to reveal the relationship between users if it is less popular in general.
We denote the number of times a particular hashtag $\hashtag$ was shared by each user $u_i \in \User$ as a hashtag-user count vector $\vec{n_{h}} = (n_{h, u_1}, n_{h, u_2}, \ldots n_{h, u_{|U|}})$, where $n_{h, u_i}$ is the number of times $u_i$ shares $\hashtag$.
We define $\mathit{x_{minH}}$ as the minimum of the total number of times all users shared a hashtag among the set of common hashtags between $u$ and $v$. We represent the total number of times all users $U$ shared a hashtag $h_i$ by $n_{h_i}= \sum_{u \in U} n_{h_i, u}$.

\begin{align}
\mathit{x_{minH}} &= min(n_{h_i} : (\hashtag_i \in \Hashtag_u \cap \Hashtag_v) )
\end{align}

Following the common Adamic Adar (Frequency-Weighted Common Neighbors) \cite{adamic2003friends} score, we define $\mathit{x_{aaH}}$ as the sum of the reciprocals of the logarithm of $n_{h_i}$ which is the number of times a hashtag $\hashtag_i$ common between $u$ and $v$ was shared by all users.
\begin{align}
\mathit{x_{aaH}} &=\sum_{\hashtag_i \in \Hashtag_u \cap \Hashtag_v} \frac{1}{\log{n_{h_i}}}
\end{align}

The Adamic Adar measure improves counting of common features by giving more weight to rarer features.  
 We  adapt the measure in an Information Theoretic approach to formalize the intuitive notion that lower entropies are more informative.
We define entropy $E$ of a hashtag $\hashtag_i$ as the entropy of its hashtag-user count vector over all users $\User$ such that $E_i = E (\vec{n_{h_{i}}})$.
Therefore, we define $\mathit{x_{minE}}$ as the minimum among the entropies of all common hashtags,
\begin{align}
\mathit{x_{minE}} &= min(E_i : (\hashtag_i \in \Hashtag_u \cap \Hashtag_v) )
\end{align}
and $\mathit{x_{aa}}$ as the sum of the reciprocals of entropies of each common hashtag 
\begin{align}
\mathit{x_{aaE}} &= \sum_{\hashtag_i \in \Hashtag_u \cap \Hashtag_v} \displaystyle\frac{1}{E_i}  
\end{align}

In the direction of the second property, we also measure the number of times different users share a particular hashtag as a count vector for a hashtag and then calculate the entropy of the count vectors for all hashtags in our dataset.
Figure \ref{fig:htperc} (right) further shows that hashtags whose usage count vectors have lower entropies are more indicative of social relations between their users. 
We therefore formalize the property that rare hashtags with lower entropies are more useful. We define $\mathit{x_{w\_aaE}}$ as $\mathit{x_{aa}}$ weighted by the number of hashtags shared by both users and $\mathit{x_{f\_aaE}}$ as $\mathit{x_{aa}}$ weighted by the sum of the reciprocals of entropies of hashtags shared by both users. 

\begin{align}
\mathit{x_{w\_aaE}} &= \displaystyle\frac{\mathit{x_{aaE}}}{|\Hashtag_u \cup  \Hashtag_v|} \\
\mathit{x_{f\_aaE}} &= \displaystyle\frac{\mathit{x_{aaE}}}{  \sum\limits_{\hashtag_i \in \Hashtag_u \cup \Hashtag_v} \displaystyle\frac{1}{E_i}}
\end{align}

Thus we represent the feature vector for the Hashtag Adversary $\HA$ for a pair of users $\pair$ as $\vec{x}^{ \Hashtag}_{\pair}=( \mathit{x_{comm}}, \mathit{x_{frac}}, \mathit{x_{dotpro}} , \mathit{x_{cosine}}, $ $ \mathit{x_{minH}}, \mathit{x_{aaH}} , \mathit{x_{minE}}, \mathit{x_{aaE}},  \mathit{x_{w\_aaE}}, \mathit{x_{f\_aaE}} )$

\subsection{Evaluation}\label{eval_ht}
We evaluate the attack in the same way as for the caption attack, i.e. using  AUC for a random forest with five-fold cross validation.
We first explain the data preprocessing, followed by a baseline method proposed in a recent work. Finally we present the results of the experiments.

\begin{table}[!ht]
\centering
\caption{Pre-processed hashtag data statistics. \label{ht_pro_dataset}}
\scalebox{0.9}{
\begin{tabular}{lcc}
\toprule
& LA & NY \\ 
\midrule
 posts & 1,712,500 & 2,999,603 \\ 
users & 10,359 &18,896  \\ 
hashtags & 406,692 & 595,502 \\ 
friend pairs & 5,229 & 12,765 \\ 
\bottomrule 
\end{tabular}
}
\end{table}

\subsubsection{Data preprocessing}
\label{ht_data}
We have 2,306,631 unique hashtags shared by 10,418 users
in our raw dataset from LA. For NY, we have 3,090,451 unique hashtags shared by 19,015 users.
In order to remove hashtags that are shared too frequently and too infrequently, we filter out hashtags that are shared by less than 2 and more than 10 users.  

Following this step, we obtain 5229 out of a total of 17,043 friend pairs from LA
and 12,765 out of 40,883 friend pairs from NY.
We randomly sample an equal number of stranger pairs.
Table \ref{ht_pro_dataset} shows the number of users, terms and friend pairs that we have after preprocessing.

\subsubsection{Baseline}
We consider as baseline, the only prior work by Zhang et al. \cite{zhang2019language} that rely on hashtags to perform friendship prediction. 
This method uses a user-hashtag bipartite graph embedding model to automatically summarize hashtag profiles as features for users.

\subsubsection{Results}
\begin{figure}[t!]
\centering
\includegraphics[width=\columnwidth]{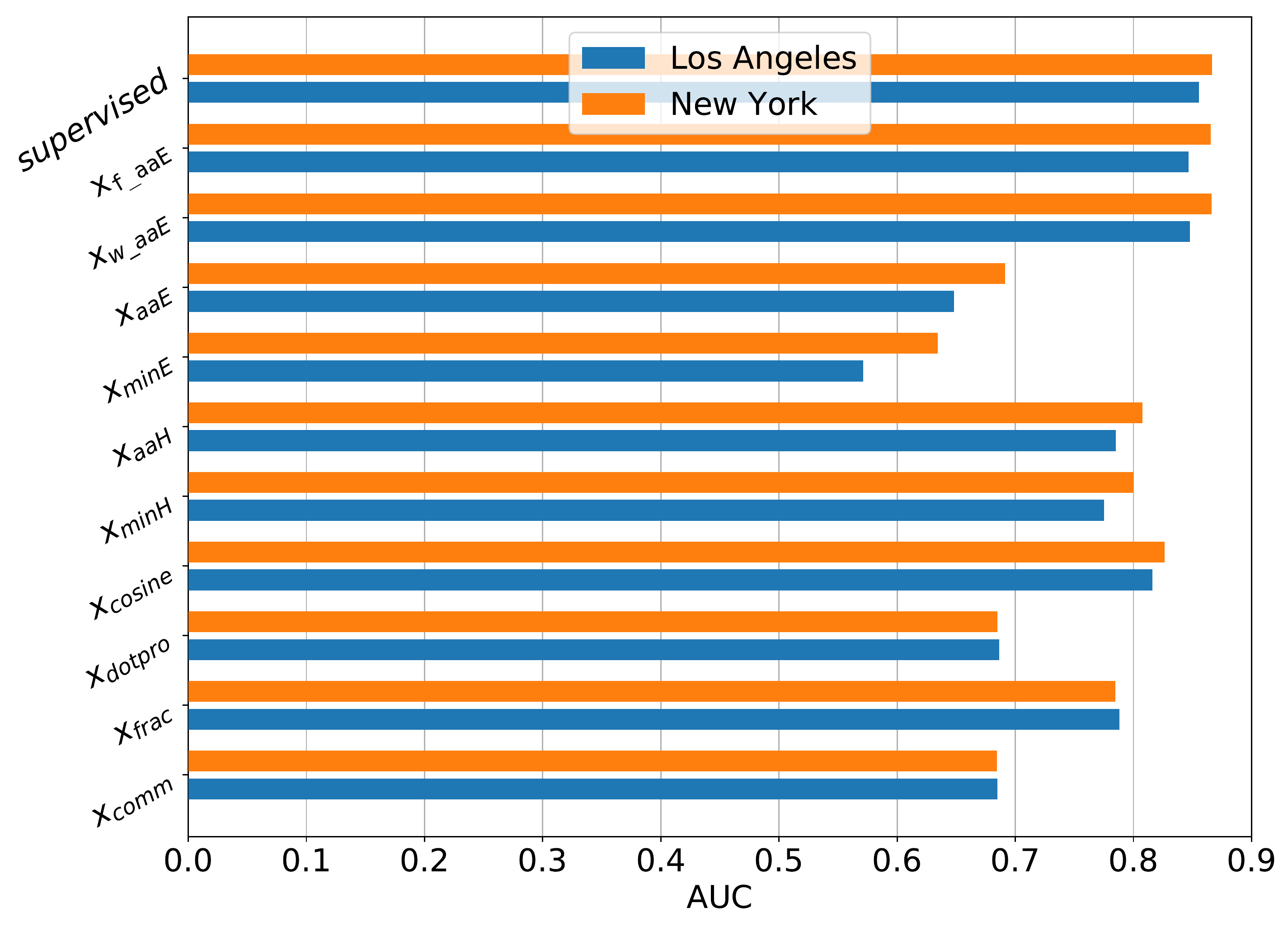}
\caption{Comparison of individual features for our Hashtag Adversary \label{fig:ht_unsup}}
\end{figure}

We first compare the performance of the 10 individual features on their own by evaluate each in an unsupervised setting. 
Figure~\ref{fig:ht_unsup} shows the AUC values.
We see that $\mathit{x_{w\_aaE}} $ achieves the highest score with an AUC of 0.87 for NY and 0.85 for LA, closely followed by $\mathit{x_{f\_aaE}}$ with an AUC of 0.87 for NY and 0.85 for LA. Both features are based on the property that rare hashtags with lower entropies are more informative.
At the third place is the cosine distance between the user-hashtag count vectors $ \mathit{x_{cosine}}$  with an AUC of $0.8263$ 
followed by $\mathit{x_{aaH}}$, $  \mathit{x_{minH}}$ and $\mathit{x_{frac}}$ with AUCs of 0.8076, 0.7999 and 0.7848 respectively.
We see that the remaining features are significantly outperformed. Nevertheless, they perform much better than random guessing.

These results show that the success of our hashtag attack is independent of the city from which we collect our data.
Although 6 out the 10 features perform sufficiently well on their own, using them in a supervised setting with the random forest yields stronger performances in both cities, with an AUC value of 0.86 for LA and 0.87 for NY. 

In contrast, the baseline method achieves an AUC of 0.8 in Los Angeles and 0.82 in New York.
Its also noteworthy that this method involves training many hyperparameters using cross validation just for learning features. 
This huge overhead in processing required in this baseline, combined with around 7\% lower AUC compared to our method, clearly demonstrates the strength of hand-engineered features for a privacy attack such as friendship inference.

%% file: images.tex
\section{Attack using Images}
\label{images}

In this section, we describe our friendship inference attack using images followed by an experimental evaluation.

\subsection{Attack Methodology}
\label{imageattack}

The Image Adversary ($\IA$) tries to infer social relationships using the content of the images posted by users on an OSN. 
We do not use facial recognition in these images following the assumption that users who wish to hide their friendship with a certain people would refrain from posting a picture with them in the first place.

\subsubsection{Extracting Information from Images.}

The recent advent of deep neural networks (NNs) and the availability of massive computing power and datasets, has facilitated huge breakthroughs in image classification. 
In addition to high image classification accuracies, pre-trained NNs can directly be used on any image to automatically obtain a meaningful feature representation of the image. 
This is why we choose a NN based approach for our Image Adversary.

We use a Residual Neural Network (ResNet) which is an advanced type of a convolutional NN for meaningfully extracting information from images.
ResNets solve the problem of degrading/saturating accuracy that is faced by increasing the depth of deep convolutional NNs \cite{He2016DeepRL}. 
Moreover, training ResNets is easier than training convolutional NNs when they need to get deeper in order to solve increasingly complex tasks. 
We adopt a ResNet namely the Places365-ResNet18 model for PyTorch trained on the Places365-Standard Database~\cite{zhou2017places} which allows learning of deep scene features for various scene recognition tasks. 
The training set consists of ~1.8 million images from 365 scene categories, with at most 5000 images per category, consistent with real-world frequencies of occurrence. 
\footnote{The scene category list can be found on \url{https://github.com/metalbubble/places_devkit/blob/master/categories_places365.txt}}

We now describe the feature design process for our attack.

\subsubsection{Feature Design.}
We extract the final softmax output layer produced by the neural network for each of the images shared by users in our dataset. 
For an image $I_\img$, the final layer is a posterior probability distribution over the 365 classes that correspond to the 365 different scene categories $c_1, \dots, c_{365}$  which we represent by $\vec{\mathit{C}_\img}= $Pr$(c_1, \dots, c_{365} | I_\img) $.
For each user $u$ we count the number of images that belong to each of the 365 scene categories. 
An image belongs to a category if its corresponding class membership probability outputted by the ResNet is higher than a certain percentile threshold (which we describe during the dataset pre-processing in \cref{im_data}).
We denote this count by the vector $\vec{n_\user}=(n_{u,1}, n_{u,2}, \ldots n_{u,365})$ where $n_{u, c}$ represents the number of images shared by user $u$ that depict the scene corresponding to category number $c$. 
We use $\vec{n_v}$ analogously for user $v$ in a pair.
We then design a set of 4 feature types for each user pair $\pair$ as follows:

1. We define the feature $\mathtt{min\_count}$ as the category-wise minimum of the 2 count vectors $\vec{n_\user}$ and $\vec{n_v}$ i.e.  out of the number of images posted by both users in each of the 365 categories, we count the minimums, $\mathtt{min\_count}= ( \mathtt{\min{(n_{u,1}, n_{v,1})},  min(n_{u,2}, n_{v,2})}, \ldots$  
$ \mathtt{\min{(n_{u,365}, n_{v,365})}} ) $

2. We define the feature $\mathtt{x_{cosine}}$ as the cosine distance between the category-wise count vectors $\vec{n_\user}$ and $\vec{n_v}$.

3. We define the feature $\mathtt{x_{F\_maxcat}}$ as the maximum over the 365 components of the vector $\mathtt{min\_count}$, which represents the frequency of the mutually most frequently shared category in images shared by both users. $\mathtt{x_{F\_maxcat}}=$ \\ $ \mathtt{ max( min(n_{u,1}, n_{v,1}),  min(n_{u,2}, n_{v,2}), \ldots }$ $ \mathtt{min(n_{u,365}, n_{v,365}) )}$

4. We perform a category-wise sum of the posterior probabilities over all images shared by each user which we represent by a usage-vector $\vec{s_c} = (  s_{c, u_1}, s_{c, u_2}, \ldots s_{c, u_{|U|}}   )$, where $s_{c, u_j}$ represents the summed up probability that a category $c$ is depicted over all images shared by $u_j$.
We define the feature $\mathtt{x_{E\_maxcat}}$ as the entropy of the usage-vector $\vec{s_{maxcat}}$ of the mutually most frequently shared category $maxcat$.

Finally, we represent the feature vector for the Image Adversary for a pair of users $\pair$ as $\vec{x}^{\Img}_{\pair}=( \mathtt{min\_count}, \mathtt{x_{cosine}}, \mathtt{x_{F\_maxcat}},$
$ \mathtt{x_{E\_maxcat}} )$.

\begin{table}[!ht]
\centering
\caption{Pre-processed image data statistics. \label{table:im_dataset}}
\begin{tabular}{lcc}
\toprule
 & LA & NY \\ 
\midrule 
 users & 5,082 & 12,575 \\ 
 images & 2,052,619 & 4,840,670 \\ 
 friend pairs &  4,262 &  18,847\\ 
\bottomrule 
\end{tabular} 
\end{table}

\subsection{Evaluation}\label{eval_im}
We evaluate the attack in the same way as for the previous attacks, i.e. using the AUC metric for a random forest classifier and assigning 80\% of each of the friend and stranger pairs for training and 20\% of each of the friend and stranger pairs for testing. 

\subsubsection{Data preprocessing} \label{im_data}
We could download 4,857,971 photos from 8,309,482 urls for our image-based attack from LA and 5,092,951 photos from 14,671,592 urls from NY. 
We feed each of these photos onto our pre-trained ResNet model and extract the final layer, i,e the probability distribution over the 365 scene categories.
We only require the significant scene categories that are predicted to be depicted in an image.
Therefore, we first remove the category probabilities lower than 5\% by setting them to 0.
We obtain 4,262 out of a total of 17,043 friend pairs from LA and 18,837 out of a total of 40,883 friend pairs from NY.
We randomly sample an equal number of stranger pairs.
Table \ref{table:im_dataset} shows the number of users, images and friend pairs that we have after preprocessing.

\subsubsection{Results}
\begin{figure}[!t]
\centering
\includegraphics[width=\columnwidth]{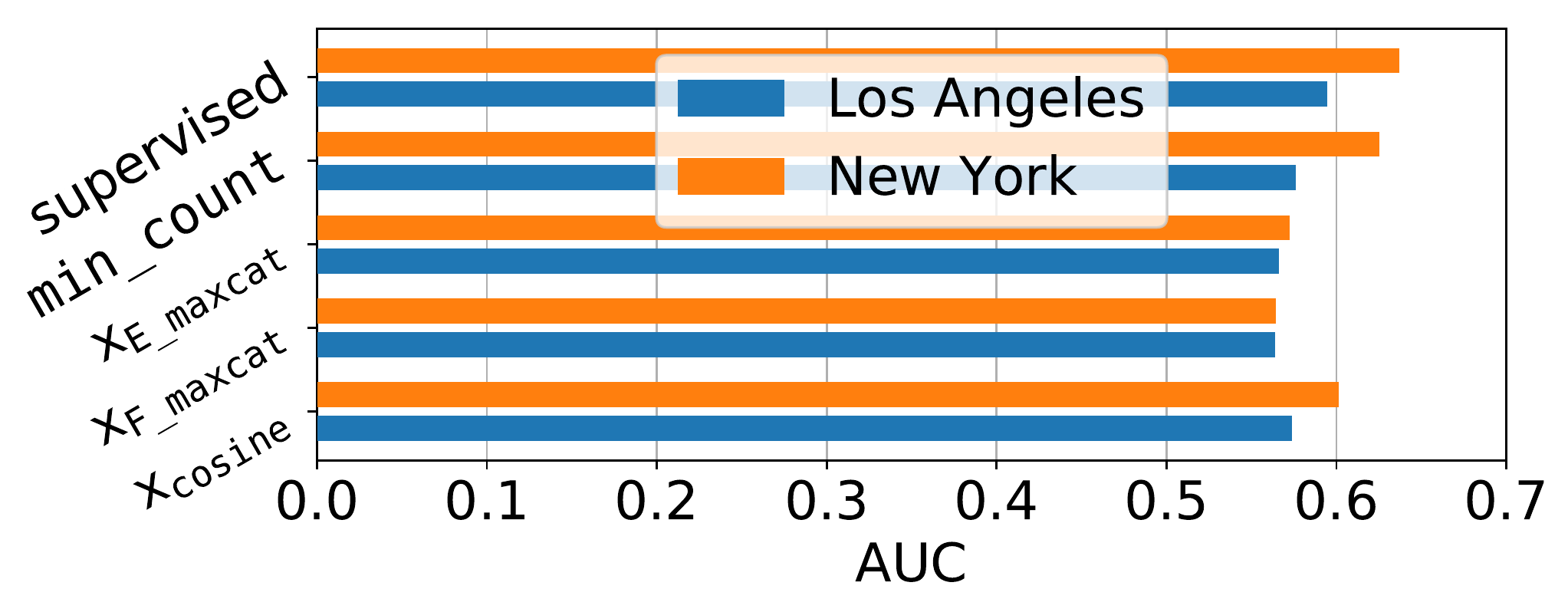}
\caption{Comparison of individual features for our Image Adversary \label{fig:im_unsup}}
\end{figure}

In order to evaluate how each of the features perform on their own, we evaluate each of them in an unsupervised setting. Figure~\ref{fig:im_unsup} shows the AUC values for the  individual features for our attack namely $\mathtt{min\_count}$, $\mathtt{x_{cosine}}, \mathtt{x_{F\_maxcat}}$ and $ \mathtt{x_{E\_maxcat}}$. 
We see that both $\mathtt{min\_count}$ and $\mathtt{x_{cosine}}$ perform equally in both cities.  
In NY, $\mathtt{min\_count}$ performs the best (AUC 0.63) followed by $\mathtt{x_{cosine}}$ (AUC 0.61).
In LA, $\mathtt{x_{cosine}}$ performs the best (AUC 0.58) followed by $\mathtt{min\_count}$ (AUC 0.57).
In the supervised setting, we achieve an AUC value of 0.64 for NY and 0.60 for LA. 
Figure~\ref{fig:sup} shows the AUC values of the image attack in the supervised setting compared with the other attacks using hashtag, text, location, partial network data as well as the multimodal attack.

Although the performance is relatively weak, we can still infer a signal and perform much better than random guessing.
Inferring the content of images is an inherently hard problem. 
Technology has not yet been able to achieve what an average human can in terms of drawing conclusions from the contents of a single image.
Most research on computer vision and face recognition techniques work only with small labelled datasets and the scalability to real world, unseen data is yet to be explored.
We are the first to perform image content analysis on a large scale real world dataset.
Since in the large scale, an adversary using computer vision techniques can perform inference attacks in a matter of seconds, privacy analysis of image content remains an important problem.

%% file: location.tex
\section{Attack using Locations and Network Information}
\label{locations}

In this section we present the details of our Location Adversary $\LA$ and Network Adversary $\NA$, both of which are based on state-of-the-art methods.
The Location Adversary ($\LA$) uses check-in data following \textit{walk2friends} and the Network Adversary $\NA$ uses already published friendships in the OSN loosely following \textit{node2vec} for edge based link prediction. 
We give a brief overview of the attack first for $\LA$ and next for $\NA$.
We follow up with experimental evaluation.

\subsection{Location Adversary}
\label{skipgram}
The Location Adversary ($\LA$) has access to locations shared by users and tries to infer whether a pair of users are socially related.
Using check-in information for inference of users' social relations has been extensively studied.
We adopt the state-of-the-art method \textit{walk2friends}\cite{walk2frns} that has been shown to outperform other approaches.

The \textit{walk2friends} method can be explained in three steps:
First, we organize users and locations into a bipartite graph,
with the number of visits by a user to a location
as the weight of the corresponding edge.
Second, multiple random walks are performed on the graph
to generate a set of walk traces with the transition probability from one node to another defined over the edge weight.
Finally, we utilize the skip-gram model~\cite{MCCD13,MSCCD13}
to embed each user into a continuous vector.
The vector for a user $\user$, is denoted by $\vec{\mathit{E}_\user}$.

To perform friendship prediction, we follow the image and text attacks above.
Concretely, for two users $\user$ and $v$,
we define the following feature vector
$
\vec{x}^{\Loc}_{\pair} = 
(\mathrm{cosine}(\vec{\mathit{E}_u},\vec{\mathit{E}_v}), \:
\mathrm{  euclidean}(\vec{\mathit{E}_u},\vec{\mathit{E}_v}),  \:
$ 
$ 
\mathrm{ correlation}(\vec{\mathit{E}_u},\vec{\mathit{E}_v}) , 
\mathrm{ chebyshev}(\vec{\mathit{E}_u},\vec{\mathit{E}_v}), \:
\mathrm{ bray\_curtis}(\vec{\mathit{E}_u},\vec{\mathit{E}_v}),  \:
$ \\ $
\mathrm{  canberra}(\vec{\mathit{E}_u},\vec{\mathit{E}_v}),
\mathrm{  manhattan}(\vec{\mathit{E}_u},\vec{\mathit{E}_v}) ,\:
\mathrm{  sq\_Euclidean}(\vec{\mathit{E}_u}, \vec{\mathit{E}_v}) 
)$ \\
and train a random forest classifier to perform our attack.

\subsection{Network Adversary}
\label{networkattack}
Our network adversary ($\NA$) has access to a limited fraction of relationships in the online social network 
and tries to predict these missing relationships.
This is inline with the traditional setting of link prediction.
We leverage state-of-the-art tool \textit{node2vec} to implement $\NA$~\cite{GL16},
which has been shown to outperform traditional methods,
such as Jaccard Index, Adamic-Adar score, and Preferential Attachment.

\textit{node2vec} first performs random walk on the partial social network,
this again results in a set of truncated random walk traces.\footnote{We follow the default parameter setting
of \textit{node2vec} in the random walk.}
Then, skip-gram is adopted to embed each user $\user$
into a vector $\vec{\mathit{F}_u}$.
For two users $\user$ and $v$,
we define their feature vector
as the Hadamard product between their vectors~\cite{GL16}, i.e.,
$\vec{x}^{\Edge}_{\pair}  = \vec{\mathit{F}_u} \circ \vec{\mathit{F}_v}$.

\begin{table}[!ht]
\centering
\caption{Pre-processed location and network data statistics.}
\begin{subtable}{0.9\columnwidth}
\centering
\caption{ \label{table:loc_dataset}}
\begin{tabular}{lcc}
\toprule
& LA & NY \\ 
\midrule 
users & 10,430 & 19,038 \\ 
locations & 23,452 & 30,073 \\ 
friend pairs & 17,043 & 40,883 \\ 
\bottomrule 
\end{tabular}
\end{subtable}
\\
\begin{subtable}{0.9\columnwidth}
\centering
\caption{\label{table:network_dataset}}
\begin{tabular}{lcc}
\toprule
& LA & NY \\ 
\midrule 
users & 9,347 & 17,853 \\ 
friend pairs & 16,294 & 39,756 \\ 
\bottomrule 
\end{tabular}
\end{subtable}
\end{table}

\subsection{Evaluation}\label{eval_loc}
We evaluate the attacks in the same way as described previously, i.e. using the AUC metric for a random forest classifier and assigning 80\% of each of the friend and stranger pairs for training and 20\% of each of the friend and stranger pairs for testing.

\subsubsection{Data Preprocessing}
\label{loc_data}
For the Location Adversary, following \cite{walk2frns}, we filter out users who have not visited at least two \emph{different} locations since such accounts with many check-ins at one location 
are most likely to be local businesses for example bars and pubs. We further filter out users with less than 20 check-ins, whom we consider to be \emph{active users}.
We obtain 17,043 friends pairs from LA and 40,883 from NY.

For the Network Adversary, we start with the friend pairs generated for the hashtag, location, image, and text based attacks as described above.
We randomly sample 80\% of these friend pairs to generate a partial network graph $\partOSN$ which we use for building the $node2vec$ embedding for users. These set of edges in our training set is a proper subset of the original graph $\OSN$.
We obtain 16,294 friends pairs from LA and 39,756 from NY.
For both attacks, we randomly select an equal number of stranger pairs out of the set of already sampled stranger pairs for the hashtag, image, and text attacks.
Tables \ref{table:loc_dataset} and \ref{table:network_dataset} present the preprocessed data statistics for both attacks.

\subsubsection{Results}
We achieve AUC values of 0.77 for both LA and NY, respectively for the location based attack.
Though different datasets, our results are quite close to the one reported in~\cite{walk2frns}.
We achieve AUC values of 0.75 and 0.78 for LA and NY respectively for the network based attacks. 

%% file: holisticattack.tex
\section{Multimodal Attack}
\label{holistic}

We now present the Multimodal Adversary $\HolA$, which leverages information from all 5 components of user posts presented in the previous sections, to infer social relationships. 
We first present the attack methodology along with a variant called the subset adversary. We follow up with an empirical evaluation as well as robustness of the attacks against information hiding.

\smallskip
\noindent\textbf{Overview of the Fusion Mechanism} The previous attacks were based on training a machine learning classifier which required a feature design process (hand-engineered or aided by deep learning techniques).  
For this attack, we use the features already designed for the 5 individual post components as described in each previous attack section.
However, we weigh the information from each of the 5 components according to their relative trustworthiness, which we infer from the strength of the individual attacks.
To this end, we divide the training set $\Pair_{\textbf{train}}$ into 2 sets.
We use one set to train the 5 random forest classifiers using the 5 respective feature vectors.
We use the second set to assign a confidence score $a$ to each of the 5 trained classifiers using their respective AUC values.
We use these AUC values as indicators of the strength or trustworthiness of the predictions made by the 5 classifiers on the target set $\Pair_{\textbf{tar}}$.

\subsection{Attack Methodology}
\label{holisticattack}
 
We describe this attack in 4 stages.

In the first stage, we divide the training set $\Pair_{\textbf{train}}$ into 2 parts:  $\Pair_{\textbf{train-train}}$  and  $\Pair_{\textbf{train-test}}$. 
We train the classifiers for the five modalities separately using the feature vectors $\vec{x}^{\Hashtag}_{\pair}, \vec{x}^{\Capt}_{\pair}, \vec{x}^{\Img}_{\pair},$ $ \vec{x}^{\Loc}_{\pair} $ and $\vec{x}^{\Edge}_{\pair}$ on the set $\Pair_{\textbf{train-train}}$.

In the second stage, from each of the 5 trained random forests, we extract the respective posterior probabilities for the friend class, i.e. the probability with which a sample pair in the set $\Pair_\textbf{tar}$ is socially related. 
In a random forest, the predicted posterior probabilities of an input sample are computed as the mean predicted class probabilities of the trees. 
The class probability of a single tree is the fraction of samples of the same class in a leaf. 

Let $x^{H}$, $x^{C}$, $x^{I}$, $x^{L}$ and $x^{E}$ denote the posterior probabilities for the friend class for a sample $\pair \in \Pair_\textbf{tar} $ as predicted by the classifier trained using hashtag, text, image, location and friendship information respectively.
Therefore we have for $\data \in \{ \Hashtag, \Capt, \Img, \Loc, \Edge \}$ ,
$x^{\data}_{\pair} = \Pr(\RelVar = 1 | \vec{x}^{\data}_{\pair})$,
where $\RelVar$ is the random variable representing the relationship between the user pair $\pair$.

In the third stage, we calculate a confidence score $a$ for each of the 5 attack components above. 
Let $a^\Hashtag$, $a^\Capt$, $a^\Img$, $a^\Loc$ and $a^E$ denote the confidence scores of the hashtag, text, image, location and friendship attacks respectively.
We use the Area under the ROC curve (AUC) score between the true class labels and the predicted posterior probabilities for the set $\Pair_{\textbf{train-test}}$ for the friend class as the confidence score. 
Therefore we have for $\data \in \{ \Hashtag, \Capt, \Img, \Loc, \Edge \}$ and $\pair \in \Pair_{\textbf{train-test}}$, 
\begin{align*}
a^\data = \mathtt{AUC}(y_{\pair}, x^{\data}_{\pair})
\end{align*}
 where $y$ denotes true class labels and $ x^{\data}_{\pair}$ denotes the predicted posterior probabilities.

In the fourth and final stage, we perform a weighted averaging of the 5 posterior probabilities for each sample $\pair \in \Pair_\textbf{tar} $ using their normalized confidence scores. We obtain the final score $s^{\HolA}_{\pair}$ for a sample $\pair \in \Pair_\textbf{tar} $ for the Multimodal Adversary $\HolA$ as,
\[
s^{\HolA}_{\pair} = \sum_{\data \in \{ \Hashtag, \Capt, \Img, \Loc, \Edge \}} x^\data_{\pair} * \frac{a^\data}{\sum\limits_{\data \in \{ \Hashtag, \Capt, \Img, \Loc, \Edge \}} a^\data} 
\]

\subsection{Subset Adversaries}
\label{sec:subset}

We further analyze how various combinations of individual post components perform towards inferring friendship between OSN user pairs. To this end we model a reduced Multimodal Adversary which uses a proper subset of the set of texts, hashtags, images, locations and friendship-edges published by OSN users, such that the adversary does not use at least 1 component out of the 5 post components. 
We call such an adversary that uses a subset $D'$ of post components, a Subset Adversary $\SA^{D'}$. 
The adversary uses any combination of a minimum of 2 upto a maximum of 4 out of the 5 post components, thus we have 25 different possible combination. 
Therefore we calculate the corresponding score $s^{D'}_{\pair}$ as:
\[
s^{D'}_{\pair} = \sum_{\data \in D'} x^\data_{\pair} * \frac{a^\data}{\sum\limits_{\data \in D'} a^\data} 
\]
where $D' \subset \{ H, C, I, L, E  \}$ and $ 2 \leq |D'| \leq 4$

\begin{table*}
\centering
\caption{Performance of the subset adversaries (multimodal attacks using size-2, size-3 and size-4 subsets of the 5 post components) for LA and NY. The best performance for subsets of each size is marked in bold. \label{subset}}
  \begin{subtable}{.3\linewidth}
  \centering
  \caption{Attacks using 2 components}
    \begin{tabular}{lcc}
    \toprule
    Attack Data & LA & NY\\ 
    \midrule
    $\{ \Capt, \Edge\} $ & 0.865  &  {\bf 0.885}  \\ 
    $\{ \Loc, \Edge\} $ &  {\bf0.909}   &  0.836 \\ 
    $\{ \Capt, \Loc\}$ &  0.901 & 0.876 \\ 
    $\{ \Hashtag, \Capt\}  $ & 0.821  & 0.842 \\ 
    $\{ \Hashtag, \Img \}  $ & 0.767   & 0.761  \\ 
    $\{ \Hashtag, \Loc\}$ &  0.805 & 0.796  \\ 
    $\{ \Hashtag, \Edge \} $ & 0.785 & 0.819 \\ 
    $\{ \Capt, \Img\} $ & 0.818  & 0.834 \\ 
    $\{ \Loc, \Img\} $ & 0.825 & 0.767 \\ 
    $\{ \Img, \Edge\} $ &  0.709  & 0.742 \\ 
    \bottomrule 
    \end{tabular} 
  \end{subtable}
  \begin{subtable}{.3\linewidth}
  \centering
  \caption{Attacks using 3 components}
    \begin{tabular}{lcc}
    \toprule
    Attack Data & LA & NY\\ 
    \midrule
     $\{ \Capt, \Edge, \Loc\}$ & {\bf0.928} &  {\bf 0.905} \\ 
    $\{ \Hashtag, \Loc, \Edge\} $  & 0.913 & 0.857  \\ 
    $\{ \Hashtag, \Capt, \Img\} $ & 0.816  & 0.835  \\ 
    $\{ \Hashtag, \Capt, \Loc\} $ & 0.903  & 0.877 \\ 
    $\{ \Hashtag, \Capt, \Edge\} $ & 0.860  &  0.883 \\ 
    $\{ \Hashtag, \Img, \Loc\}  $ & 0.843  & 0.800 \\ 
    $\{ \Hashtag, \Img, \Edge\}$  & 0.758  & 0.791\\ 
    $\{ \Capt, \Img, \Edge\} $ & 0.859 & 0.879\\ 
    $\{ \Capt, \Img, \Loc\} $  &  0.901  & 0.875 \\ 
    $\{  \Img, \Loc, \Edge\}  $ & 0.903  &  0.830 \\ 
    \bottomrule 
    \end{tabular} 
  \end{subtable}
 \begin{subtable}{0.3\linewidth}
 \centering
    \caption{Attacks using 4 components}
    \begin{tabular}{lcc}
    \toprule
    Attack Data & LA & NY\\ 
    \midrule
      $\{ \Hashtag, \Capt, \Loc, \Edge\} $& {\bf 0.927}  &  {\bf 0.905}\\ 
      $\{ \Capt, \Img, \Loc, \Edge \} $ & {\bf0.927}  & 0.902 \\ 
		$\{ \Hashtag, \Capt, \Img, \Loc\}$ & 0.903 &   0.877 \\ 
      $\{ \Hashtag, \Capt, \Img, \Edge\} $&  0.854 & 0.877  \\ 
      $\{ \Hashtag, \Img, \Loc, \Edge\}$ & 0.908 & 0.852\\ 
    \bottomrule 
    \end{tabular} 
  \end{subtable}
\end{table*}

\subsection{Evaluation}\label{eval_hol}
We first present a baseline approach which we design to evaluate the multimodal attack. We then describe the dataset and evaluation metric.
We then discuss the results, followed by an analysis of the interplay between the different modalities.
We end the section with an analysis of the robustness of the attack against the amount of data present in the OSN.

\subsubsection{Baseline}
 We implement a baseline that performs a simple averaging of the posterior probabilities from each data component. Thus the baseline score $s^{BL}_{\pair}$ for a sample $\pair \in \Pair_\textbf{tar} $ with posterior probabilities $x^{\data}_{\pair}$ where $\data$ denotes the  individual data component which is used by each attack is defined as
$s^{BL}_{\pair} = \frac{\sum_{\data \in \{ \Hashtag, \Capt, \Img, \Loc, \Edge  \}}
 x^\data_{\pair} }{|D|}$

\subsubsection{Dataset and metric}
We use the same dataset as for the individual attacks. 
As mentioned earlier, we obtained 17,043 friend pairs for LA and 40,883 friend pairs for New York. We collect the stranger pairs sampled in the hashtag, image, and textual attacks. Each pair in this resultant dataset for the Multimodal Adversary has feature values for at least one out of the set $\{ \vec{x}^{\Hashtag}_{\pair}$ (hashtag), $\vec{x}^{\Img}_{\pair}$ (image), $\vec{x}^{\Capt}_{\pair}$ (text), $\vec{x}^{\Loc}_{\pair}$ (location) and $\vec{x}^{\Edge}_{\pair}$ (partial network) $\}$.

Following the other attacks, we evaluate this attack also using the AUC metric.
However, unlike the other attacks for single post components, for the multimodal attack, we only calculate a score $s_{\pair}$ for each sample pair instead of designing a set of features. 
Thus we do not need to employ a supervised machine learning classifier, instead we directly calculate the area under the ROC curve for the multimodal scores $s_{\pair}$ of the target pairs.

\begin{figure}[t!]
\centering
    \includegraphics[width=\columnwidth]{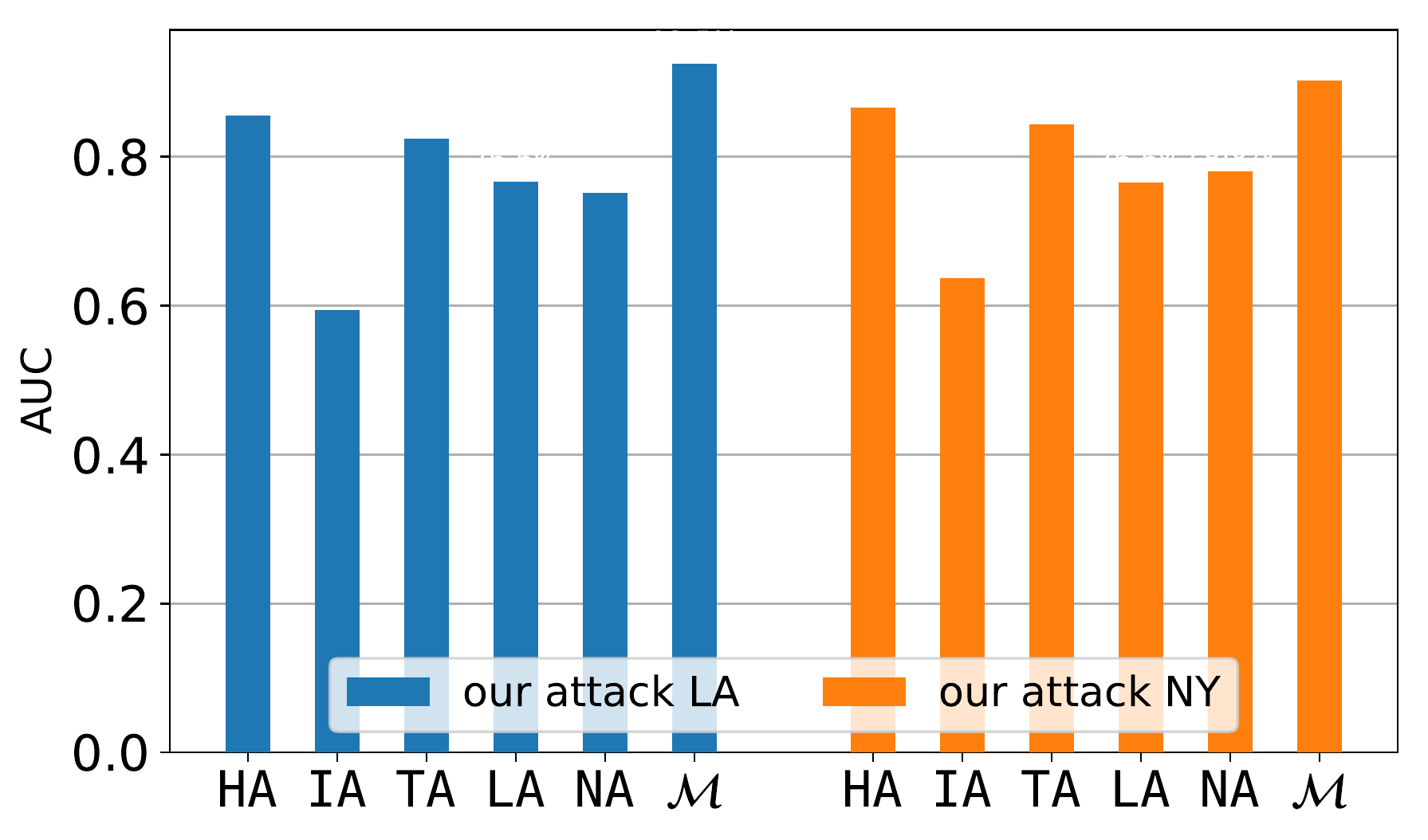}

\caption{Comparison of the AUC values achieved by all 6 adversaries (the Hashtag ($\HA$), Image ($\IA$), Text ($\CA$), Location ($\LA$), Network ($\NA$) and the Multimodal Adversary $\HolA$).
\label{fig:sup}}
\end{figure}

\subsubsection{Results}
We now discuss the results of our multimodal and subset adversaries.

\smallskip
\noindent\textbf{Multimodal Adversary $\HolA$.} For our Multimodal Adversary using the weighted average score $s^{\HolA}_{\pair}$ as described above, we achieve a very strong performance in both cities with an AUC of 0.925 for LA and 0.905 for New York. 
For the baseline approach, using the simple average score $s^{BL}_{\pair}$, we achieved an AUC of 0.924 for LA and 0.875 for New York 

We also compare the multimodal attack with the 5 monomodal attacks in Figure~\ref{fig:sup}.

The multimodal attack achieves the strongest performance for both cities. Moreover, the attacks using only hashtags and only text significantly outperform the attacks using only images, only locations and only friendship edges. 
We also list these AUC values in Table \ref{subset} for comparison with the subset adversaries.

\smallskip
\noindent\textbf{Subset Adversaries $\SA^{D'}$.}
Table \ref{subset} presents the AUCs of size-2, size-3 and size-4 subset adversaries. The best performance for subsets of each size is marked in bold.
Among all subset sizes, the attack using the size-3 set of texts, friendship-edges, and location information performs the best for both LA and New York with AUCs of 0.928 and 0.905 respectively. 
The size-4 set of texts, locations, edge-friendships and hashtags perform equally strongly for New York with the same AUC and only slightly less strongly for LA with an AUC of 0.927. 
The highest performance among size-2 subsets is achieved by the combination of $\LA$ and $\NA$ (0.909) for Los Angeles and the combination of $\CA$ and $\NA$ (0.885) for New York.

The worst performance in both cities is seen in the attack combining images and friendship-edges with an AUC of 0.709 for LA and 0.742 for New York.  This result is not surprising given that the individual attacks using only image have the worst performance for both cities and for LA, the attack using only friendship-edges is the second worst. For New York it is the attack using only locations that is the second worst and unsurprisingly the set using both images and locations has one of the weakest AUCs of 0.767.

\subsubsection{Interplay between Modalities}
We now discuss some interesting observations on the effect of different modalities on each other.

\begin{figure}[t!]
\centering
\includegraphics[width=\columnwidth]{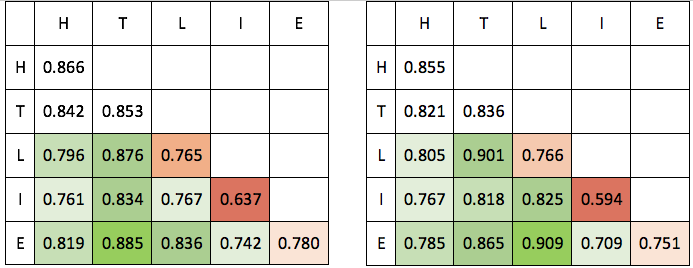}
\caption{Matrix showing AUCs achieved by complementary components (size-2 subset adversary) left: NY, right: LA. \label{matrix}}
\end{figure}

\smallskip
\noindent\textbf{Complementary Components.} The diagonal elements in Figure \ref{matrix} further show the AUCs of the monomodal adversaries and the non-diagonal elements show size-2 subset adversaries. The greener elements show increasing AUCs and the redder elements show decreasing AUCs.
We see from Figure \ref{matrix} that certain components complement each other e.g. for the LA dataset, locations and images on their own achieve very low AUCs of 0.7663 and 0.594 respectively but when combined together the AUC goes up to 0.825 which is a significant increase! Edges achieve an AUC of 0.751 on its own, but when it is combined with locations, the AUC shoots to 0.909! Yet another example is text which achieves an AUC of 0.836 on its own but together with locations achieves a much higher AUC of 0.901! These examples demonstrate that our fusion mechanism for the Multimodal Adversary exploits these complementing components.

\smallskip
\noindent\textbf{Combination of Weaker Components,}
Firstly, we note that the best performing size-4 subset but with images instead of hashtags ($D'=\{ \Capt, \Img, \Loc, \Edge\} $) achieves the same AUC in LA and only slightly lower AUC of 0.902 for New York.
This result is very interesting given that our independent attack using images only is much weaker than our attack using hashtags only. 
Moreover, both the attacks using hashtags only and texts only, significantly outperformed the attacks using only images, only locations and only friendship edges.
Thus it is very interesting to see that the attack using images, locations and friendship-edges is significantly stronger than the attack using hashtags together with texts (with an AUC of 0.903 for the former versus 0.821 for the latter).  The performance of images, locations and friendship edges improves tremendously even when combined with only 1 additional component.
Also surprising is the attack using images along with locations for LA. It is stronger than the attack using images along with both hashtags and friendship-edges (with an AUC of 0.825 for the  former versus 0.758 for the latter). This is interesting because the attack using only locations has one of the poorest performance for LA.
Thus we deduce that even though some components are clearly outperformed by others when used individually, they may contribute significantly towards a much stronger inference attack when combined together with other weak components.

\smallskip
\noindent\textbf{ Increasing Number of Components, }
In a majority of the cases, we see from Table \ref{subset} that combining additional components improves the attack success. For example adding one component to either of the size-3 subsets  $\{ \Hashtag, \Capt, \Img\} $,  $\{ \Hashtag, \Img, \Loc\} $, $ \{ \Capt, \Img, \Loc\}$ and $\{ \Hashtag, \Capt, \Loc\}$ to get the size-4 subset $\{ \Hashtag, \Capt, \Img, \Loc\}$ always improved the AUC. The same can be observed for other size-3 to size-4 transitions such as  $\{ \Hashtag, \Capt, \Loc\} $,  $\{ \Hashtag, \Capt, \Edge\} $ or  $\{ \Hashtag, \Loc, \Edge\} $ to get $\{ \Hashtag, \Capt, \Loc, \Edge\} $ as well as size-2 to size-3 transitions such as  $\{ \Capt, \Edge\}$, $\{ \Capt, \Loc\}$ and $\{ \Edge, \Loc\}$ to $\{ \Capt, \Edge, \Loc\}$.  This underlines the effectiveness of our proposed fusion mechanism for the Multimodal Adversary. Further, the success of the attacks continue to increase consistently across cities as well. 

Thus, increasing the number of components increases the attack success for almost all cases except the transitions from the strongest subset overall, i.e. $\{ \Capt, \Edge, \Loc\}$.
Moreover, the Multimodal Adversary $\HolA$ using all 5 modalities does not have a significant increase in performance over the best performing size-3 or the size-4 subsets adversaries. 
This can be explained by the impact of the relatively weak image component, which is unable to contribute to the inference attack anymore beyond the performance achieved by the remaining components.

%

We summarize our key observations as follows:
\begin{enumerate}
\item Various pairs of components complement each others' attack success when combined.
\item A combination of weaker components often increased the attack performance more than a combination of stronger components.
\item Increasing the number of components almost always increase the attack success.
\end{enumerate}

\subsection{Robustness}
So far we have seen the severity of privacy risks arising from sharing various types of posts on an OSN. 
An immediate question that arises at this point is: Is the problem solved when users simply shared posts less frequently, how is the performance is affected if users shared a lower number of posts.
We therefore evaluate the robustness of the four attacks namely the attacks using hashtags, texts, images and a combination of all three of them in the multimodal setting. 
Robustness of the Location and Network Adversaries has been extensively studied in prior work that proposed \textit{walk2friends} and \textit{node2vec}.

\begin{figure}[t!]
\centering
\includegraphics[width=\columnwidth]{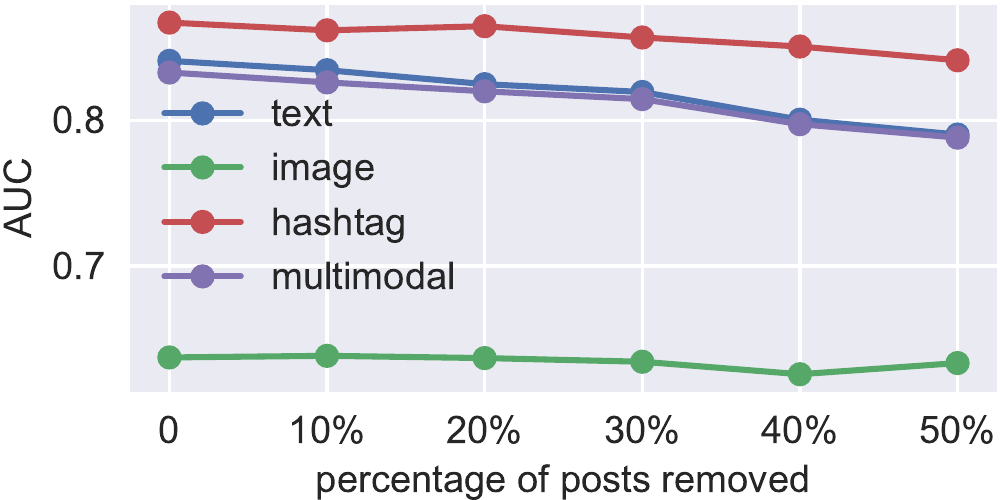}
\caption{Robustness of our attacks in the NY dataset against removing posts.}
\label{fig_defense}
\end{figure} 

We first randomly remove a certain proportion of posts published by users and collect the images, hashtags, texts contained in the remaining posts. 
Then we recompute the corresponding feature values for the training set of user pairs with which we then train three random forest classifiers, one each for hashtags, texts and images. 
We also use these three feature sets to carry out the multimodal attack on the subset $D'=\{H, C, I\}$ as described in \cref{subset}.  

We begin by removing 10\% of the original number of posts in either dataset and increase the amount of posts to be removed in steps of 10\% until 50\% of the original number of posts are removed.
The selection of posts to be removed is random.

Figure \ref{fig_defense} shows the robustness of the attacks against removal for NY. The y-axis shows the AUC attained at various percentage of posts removed shown in the x-axis. Results for LA follow a similar trend and can be found in the appendix in Figure \ref{fig_defense_LA}.
We see that the performance drops gradually for all attacks except the attack using images only. However the performance does not degrade significantly even with 50\% of the posts removed from the dataset. This shows that the attacks are quite robust to the amount of data present in the OSN. 
This calls for advanced defense mechanisms against adversaries that utilize multiple kinds of user information on OSNs for preserving both privacy as well as utility.

%% file: discussion.tex
\section{Discussion and Future Work}
\label{discussion}
Our main goal is to show whether a combination of multiple modalities improves the strength of an inference attack. 
The three individual monomodal attacks, i.e., text, images, and hashtags, presented in this paper are first of its kind,
and use the most popular and standard techniques.
However, they may be further optimized using newer techniques. 
Secondly, some of the unsupervised methods work almost as good as the supervised attacks, which demonstrates the power of the features we proposed.
Thirdly, our attack may results in false positives in case of users who are not socially related but post about similar, popular or trending topics or events. However, our attacks eliminate this effect, by favoring hashtags with low entropy and using TF-IDF for text.

We demonstrated that our attacks are robust to the amount of information shared by users. This highlights potential future work in analysis of the attack performance by obfuscating components or posts whose components have the highest contribution in predicting friendship. 
A recent work AttriGuard \cite{jia2018attriguard} takes an adversarial example based approach towards classification evasion to defend against users location (city) inference. However, these defenses are only on the personal attribute level and uses only type of public data (user's app rating scores in case of AttriGuard), whereas our attack deals with multimodal data shared by the user (thousands of images, captions, hashtags, locations). Applicability of evasion or poisoning attacks requires further research.
Secondly, we demonstrated that the privacy threat is equally severe in datasets from two different cities. The same framework can be used to analyse data collected from other OSNs or other cities all around the world.
Finally, we only provide a lower bound on the performance of the attacks. Considering factors like the date or time when 2 users shared certain posts can potentially make the attacker stronger.

%% file: related_work.tex
\section{Related Work}
\label{sec:relwork}

We present literature review in each of the 5 components of OSN data that we consider namely, link prediction, locations, text, images and hashtags followed by research on OSN data in general.
We also report existing works that analyse their implications on the privacy of users, in particular social relationships wherever applicable.

\subsection{Link Prediction} Liben-Nowell et al. \cite{liben2007link} were among the earliest to formalize the link prediction problem in social networks. 
Since then, link prediction has been extensively studied \cite{GOYAL201878}, \cite{Pecli2018}, \cite{wang2015link}, \cite{lu2011link}, \cite{Getoor}, 
including the winners of the 2011 Kaggle Social Network Challenge \cite{kaggle} that use link prediction for de-anonymization on Flickr data. We also have a large-scale real-life dataset, however 
we exploit an advanced deep-learning based approach node2vec \cite{GL16} which is shown to outperform other approaches.
Other lines of work use social links for attribute inference such as Gong et al. \cite{gong2016you} infer attributes such as major, employer, and cities lived from friendship and behavioural links. 
We, however, try to infer social links from 5 types of user generated content.

\subsection{Location} Early works such as \cite{eagle2009inferring}, \cite{crandall2010inferring}, \cite{pham2013ebm}, \cite{scellato2011exploiting}, \cite{wang2014pgt} infer social ties from co-occurrence in time and space and require two users to share common friends or locations.
Olteanu et al.~\cite{olteanu2017quantifying} study the effect of co-location information on location privacy.
More recently, Zhou et al. \cite{zhou2018vec2link} infer social links from friendships and mobility data.
Finally, Backes et al. \cite{walk2frns} take a deep learning approach to learn users' mobility features and use them for social relationship inference. 
We adopt this approach for our location based attack since it best fits our user and adversary model, and is shown to significantly outperform prior approaches \cite{pham2013ebm}, \cite{scellato2011exploiting}, \cite{wang2014pgt}.

\subsection{Text}  The growth of textual data in OSNs has been exploited by a plethora of advances in NLP and text mining.
Kim et al. \cite{kim2016extending} study the influence of offline friendship on tweeting behavior,
Adamic et al. \cite{adamic2003friends} proposed a matchmaking algorithm that leverages text, mailing list, and in and out-link information to analyze the structure of a social network.
Preotiuc-Pietro et al. \cite{PreotiucPietro2017BeyondBL} study use language features like unigrams, word clusters and emotions to predict political ideologies of users.
Tan et al. \cite{tan} characterize relations between ideas in texts using co-occurrence within documents and prevalence correlation over time.
However friendship inference from textual content has not yet been studied to the best of our knowledge.

\subsection{Images} Shoshitaishvili et al. \cite{shoshitaishvili2015portrait} combine facial recognition, and spatial analysis to determine user pairs that are dating amongst people represented in a large corpus of pictures shared on a social network.
Ilia et al. \cite{Ilia} propose an access control mechanism that allows users to manage the visibility of their face in photos published by other users. Research on defenses also mainly focuses on obfuscation of faces such as person image generation \cite{ma2018disentangled}, face replacement \cite{sun2018hybrid}, redaction \cite{orekondy2018connecting} etc. Recent work by Orekondy et al. \cite{orekondy2018connecting} focuses on localized segments containing text-like attributes such as license plate, datetime, name, etc . However these approaches are tested only on small self-annotated datasets and their applicability to friendship inference in large scale real life datasets is yet to be explored.

\subsection{Hashtags} Highfield and Leaver \cite{highfield2014methodology} examine research challenges of working with hashtags in Instagram in comparison to Twitter.
Han et al. \cite{Han} use Instagram and hashtag data to identify teens and adult age groups via behavioral differences between them. 
Using the hashtag \#selfie, authors of \cite{SCFYCQA15} study the collective behavior of sharing selfies on Instagram and \cite{LBVAASK17} study the potential dangers of self portraits. 
Authors of \cite{MAH16} study the association between emotion and health with the hashtag \#foodporn.
Prior work on privacy implications of hashtags studies location privacy using hashtags and friendship inference \cite{tagvisor}. A recent work \cite{zhang2019language} uses a user-hashtag bipartite graph
embedding model to infer friendships, but are hugely outperformed by our method.

\subsection{OSNs and Privacy}

There has been extensive research on OSN user deanonymization, based on the network topology \cite{narayanan2009anonymizing} or using metadata of tweets \cite{perez2018you}. However the adversary needs to already have access to the true identity of the target users alongwith metadata of their older tweets for the latter work.
Works on cross-site profile matching such as \cite{goga2015reliability, liu2014hydra} utilize only user profile metadata such as name, profile picture, home location etc. 
In contrast, we focus on user generated content, i.e. the multimedia footprint of users built up over time from the data they post everyday, since this collectively gives a faithful reflection of the user's interests. 
Liu et al. \cite{liu2014hydra} do consider tweets posted over time as well as friendships but they perform cross platform linkability. 
We address a different problem, i.e., friendship inference and additionally include hashtags, images, location check-ins as well. 

%% file: conclusion.tex
\section{Conclusion}
\label{sec:conclu}
We present novel attacks on the privacy of social relationships using hashtags, images and caption data. 
We also evaluate state-of-the-art tools to infer friendships using locations and friendships (between other pairs) shared on OSNs.
We finally focus on adversaries that use not just one type of user information but combine all available types of information for the inference attack. 
We demonstrate that the success of the friendship inference attacks are greatly amplified when multiple modes are combined, for example in the LA dataset, our multimodal attack achieves an AUC of 0.92 whereas the maximum AUC achieved by any of our monomodal attacks is only 0.8555 (using hashtags). 
Our experiments on the subset adversaries demonstrate that although some components clearly outperform others when used individually, they contribute towards a much stronger inference when combined together with additional components. The fusion mechanism for our multimodal adversary is capable of exploiting post components that complement each other.

We demonstrate that our attacks are robust to the amount of information shared by users. As newer data types or trends will emerge in the world of OSNs in future, the multimedia footprints of users would get increasingly diverse. The threat to user privacy arising from such multimodal adversaries would potentially be much larger. Development of effective defense mechanisms or privacy advisors is therefore the need of the hour.

%% file: appendix.tex
\section{Appendix}

\begin{figure}[!ht]
\centering
\includegraphics[width=\columnwidth]{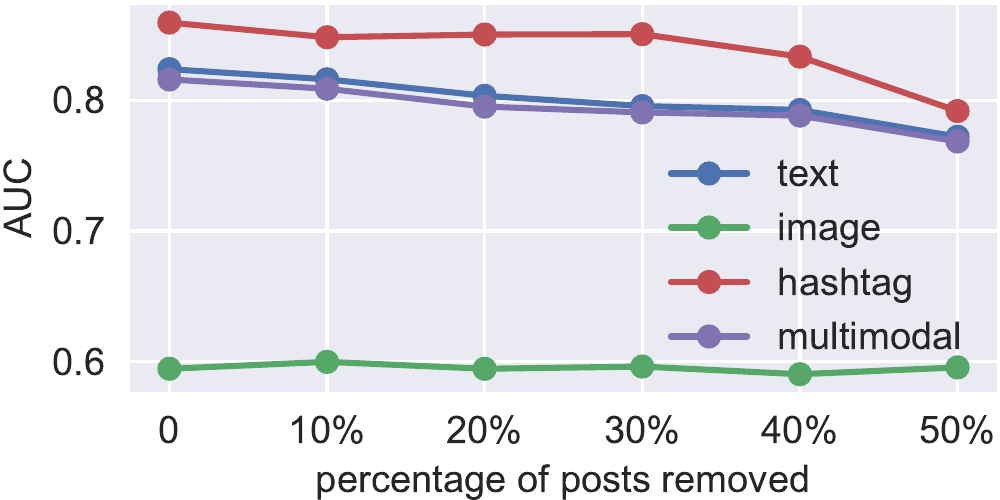}
\caption{Robustness of our attacks in the LA dataset.
The y-axis shows the AUC attained at various percentage of posts removed shown in the x-axis}
\label{fig_defense_LA}
\end{figure}